\documentclass[11pt]{article}

\usepackage{amsmath,amsthm,verbatim,amssymb,amsfonts,amscd, graphicx, xcolor, hyperref}
\usepackage{caption}
\usepackage{subcaption}
\usepackage{bm}
\usepackage{lscape}
\usepackage[color=yellow, disable]{todonotes}%
\usepackage{colortbl,hhline}
\usepackage{array,multirow,graphicx}
\usepackage[titletoc]{appendix}
\usepackage{caption}
\usepackage{colortbl,hhline}
\usepackage{array,multirow,graphicx}
\usepackage{subcaption}
\usepackage{bm}
\usepackage{lscape}
\usepackage{authblk}
\newcommand{\note}[1]{\todo[inline]{#1}}

\newcommand{\thickhline}{%
    \noalign {\ifnum 0=`}\fi \hrule height 1pt
    \futurelet \reserved@a \@xhline
}
\DeclareMathOperator*{\argmin}{arg\,min}

\begin{document}

\title{Accurate estimation of influenza epidemics using Google search data via ARGO}

\author[1]{Shihao Yang\thanks{shihaoyang@g.harvard.edu}}
\author[2,3]{Mauricio Santillana\thanks{msantill@fas.harvard.edu; corresponding author}}
\author[1]{S. C. Kou\thanks{kou@stat.harvard.edu; corresponding author}}

\affil[1]{Department of Statistics, Harvard University, Cambridge, MA, USA}
\affil[2]{School of Engineering and Applied Sciences, Harvard University, Cambridge, MA, USA}
\affil[3]{Boston Children's Hospital Informatics Program, Boston, MA, USA}




\maketitle

\begin{abstract}
Accurate real-time tracking of influenza outbreaks helps public health officials make timely and meaningful decisions that could save lives. We propose an influenza tracking model, ARGO (AutoRegression with GOogle search data), that uses publicly available online search data. In addition to having a rigorous statistical foundation, ARGO outperforms all previously available Google-search–based tracking models, including the latest version of Google Flu Trends, even though it uses only low-quality search data as input from publicly available Google Trends and Google Correlate websites. ARGO not only incorporates the seasonality in influenza epidemics but also captures changes in people’s online search behavior over time. ARGO is also flexible, self-correcting, robust, and scalable, making it a potentially powerful tool that can be used for real-time tracking of other social events at multiple temporal and spatial resolutions.

{\color{blue}This is the preprint of the paper published at PNAS: dx.doi.org/10.1073/pnas.1515373112. There are some minor differences between this preprint and the published paper.}
\end{abstract}


Big data sets are constantly generated nowadays as the activities of millions of users are collected from internet-based services. Numerous 
studies have suggested great potential of these big data sets to detect/manage epidemic outbreaks 
(influenza \cite{Ginsberg_2009, Polgreen_2008, yuan2013monitoring, Paul_etal_14, McIver_2014, Santillana15112014}, Ebola \cite{wesolowski2014}, 
dengue \cite{chan2011using}), predict changes in stock prices \cite{preis2013, bollen2011} 
and housing prices \cite{wu2014}, 
etc. In 2009, Google Flu Trends (GFT), a digital disease detection system that uses the volume of selected Google search terms to estimate current influenza-like illnesses (ILI) activity, was identified by many as a good example of how {big data} would transform traditional statistical predictive analysis \cite{Helft_2008}. 
However, significant discrepancies between GFT's flu estimates and those measured by the Centers for Disease Control (CDC) in subsequent years
led to considerable doubt about the value of digital disease detection systems \cite{butler2013}. While multiple 
articles have identified methodological flaws in GFT's original algorithm \cite{Cook_etal_11, Lazer_etal_14,Santillana_etal_14} and 
have led to incremental improvements \cite{Cook_etal_11, Santillana_etal_14, Stefansen_2014}, a statistical framework 
that is theoretically sound and capable of accurate estimation is still lacking. Here we present such a 
framework that culminates in a new method that outperforms all existing methodologies for tracking influenza activity
using internet search data.

Influenza outbreaks cause up to 500,000 deaths a year worldwide, and an estimated 3,000 to 50,000 deaths a year in the USA \cite{WHO_14}. 
Our ability to effectively prepare for and respond to these outbreaks heavily relies on the availability of accurate real-time estimates of their activity. 
Existing methods to predict the timing, duration and magnitude of flu outbreaks remain limited \cite{Shaman_2012}. Well-established clinical methods to track flu activity, such as the CDC's ILINet, report {\color{black} the percentage of patients seeking medical attention with ILI symptoms} (www.cdc.gov/flu/).
{\color{black} While CDC's \%ILI is only a proxy of the flu activity in the population, it can help officials allocate resources in preparation for potential surges 
of patient visits to hospital facilities. See \cite{lipsitch_2011, nsoesie2014systematic, chretien2014influenza} for further discussion.}

CDC's ILI reports have a delay of one to three weeks due to the time for processing and aggregating clinical information. This time lag is 
far from optimal for decision-making purposes. In order to alleviate this information gap, multiple methods combining climate, demographic and epidemiological 
data with mathematical models have been proposed for real-time estimation
of flu activity \cite{Shaman_2012, chretien2014influenza, Nsoesie_2013, soebiyanto2010modeling, shaman2013real, Yang03032015}. In recent years, 
methods that harness internet-based information have also been proposed, such as Google \cite{Ginsberg_2009}, Yahoo \cite{Polgreen_2008}, 
and Baidu \cite{yuan2013monitoring} internet searches, Twitter posts \cite{Paul_etal_14}, Wikipedia article views \cite{McIver_2014}, 
clinicians' queries \cite{Santillana15112014}, and crowd sourced self-reporting mobile apps such as Influenzanet (Europe) \cite{paolotti2014}, 
Flutracking (Australia) \cite{dalton2009flutracking}, and Flu Near You (USA) \cite{Smolinski2015}. 
Among them, GFT has received most attention and has inspired subsequent digital disease detection 
systems \cite{yuan2013monitoring, chan2011using, althouse2011prediction, ocampo2013using, scarpino2012optimizing, Davidson2015}. %
Interestingly, Google has never made their raw data public, thus, making it impossible to reproduce the exact results of GFT. 

We highlight three limitations of the original GFT algorithm, previously identified in \cite{Lazer_etal_14, Santillana_etal_14}. First, 
it was shown that a static approach, which does not take advantage of newly available CDC's ILI activity reports as the flu season evolves, 
produced model drift, leading to inaccurate estimates. Second, the idea of aggregating the multiple query terms (the independent variables in the GFT model) 
into a single variable did not allow for changes in people's internet search behavior over time (and thus changes in query terms' abilities to track flu) to 
be appropriately captured. Third, GFT ignored the intrinsic time series properties, such as seasonality of the historical ILI activity, 
thus overlooking potentially crucial information that could help produce accurate real time ILI activity estimates. 

\subsection{Our contribution}  The new methodology presented here produces robust and highly accurate ILI activity level estimates by addressing the three 
aforementioned shortcomings of the multiple GFT engines. {\color{black} In addition, we provide a theoretical framework that, for the first time, 
justifies the prevailing usage of linear models in the digital disease detection literature by incorporating causality arguments through a hidden Markov model. 
This theoretical framework contains as a special case the model developed in \cite{Santillana_etal_14}. } Our new model not only achieves the goal of (a) 
dynamically incorporating new information from CDC reports as they become available and (b) automatically selecting the most useful Google search queries 
for estimation as in \cite{Santillana_etal_14}, but also largely improves estimation by (c) including the long-term cyclic information (seasonality) 
from past flu seasons on record as input variables, and (d) using a two-year moving window (which immediately precedes the desired date of estimation) for 
the training period to capture the most recent changes in people's search patterns {\color{black}  and time series behavior \cite{Burkom_etal_2007}}. 
{\color{black} Our methodology efficiently builds a prediction model from individual search frequency as well as the past records of ILI 
activity. It utilizes both sources of information more efficiently than simply combining GFT with autoregressive terms as suggested 
in \cite{Lazer_etal_14}, since GFT is not optimally aggregated to provide additional information on top of time series information. Furthermore, we 
provide a quantitative efficiency metric that measures the statistical significance of the improvement of our methodology over other alternatives.
For example, our method is twice as accurate as the method that combines GFT with autoregressive terms (see Table \ref{tab:relative_efficiency}).} 
Finally, even though we use as input only the publicly available, low-quality data 
from the \textit{Google Correlate} and \textit{Google Trends} websites, our method has significant improvement over the latest version of GFT.

We name our model ARGO, which stands for AutoRegression with GOogle search data. Statistically speaking, ARGO is an autoregressive model with Google 
search queries as exogenous variables; ARGO also employs $L_1$ (and potentially $L_2$) regularization in order  
to achieve automatic selection of the most relevant information. 

\section{Results}
Retrospective estimates of influenza activity (ILI activity level, as reported by the CDC) were produced using our model, ARGO, for the time period of 
2009-03-29 to {\color{black}2015-07-11}, assuming we had access only to the historical CDC's ILI reports up to the previous 
week of estimation. We compared ARGO's estimates with the ground truth: the CDC-reported weighted ILI activity level, published typically with one 
or two weeks delay, by calculating a collection of accuracy metrics described in the materials section. These metrics include the Root Mean Squared Error (RMSE), 
Mean Absolute Error (MAE), {\color{black} Mean Absolute Percentage Error (MAPE),} Correlation with estimation target, and 
Correlation of increment with estimation target. For comparison, we calculated these accuracy metrics for (a) GFT estimates 
(accessed on {\color{black} 2015-07-11}), (b) estimates produced using the method of Santillana et al.\ 2014 \cite{Santillana15112014, Santillana_etal_14}
, {\color{black} (c) estimates produced by combining GFT with an AR(3) autoregressive model \cite{Lazer_etal_14}}, (d) estimates produced 
with an AR(3) autoregressive model \cite{Paul_etal_14, Lazer_etal_14}, and (e) a naive method that simply uses the value of the prior week's 
CDC's ILI activity level as the estimate for the current one. {\color{black}For fair comparison, all benchmark models (b -- d) are dynamically trained 
with a two-year moving window.}

\begin{figure}[htbp]
\centering
\caption{Estimation results. The top panel shows the estimated ILI activity level from ARGO (thick red), 
contrasting to the true CDC's ILI activity level (thick black) as well as the estimates from GFT (green), 
method of \cite{Santillana_etal_14} (blue), {\color{black}GFT plus AR(3) model (dark yellow)} and AR(3) model (dashed grey). 
The two background shades, white and yellow, reflect two data sources, \textit{Google Correlate} and \textit{Google Trends}, respectively.
{\color{black}The dashed yellow vertical line separates \textit{Google Correlate} data with search terms identified on 2009-03-28 and 2010-05-22.}
The second panel shows the estimation error, defined as estimated value minus the CDC's ILI activity level. 
The small panels labeled in alphabetical order are zoomed-in plots for estimation results in different study periods.  
Panel (a) is the H1N1 flu outbreak period. Panel (b) is the 2012-13 regular flu season. Panel (c) is the 2014-15 regular flu season. 
A regular flu season is defined as week 40 of one year to week 20 of the following year.}
\label{fig:all_pred}
\includegraphics[scale=0.5, page=1]{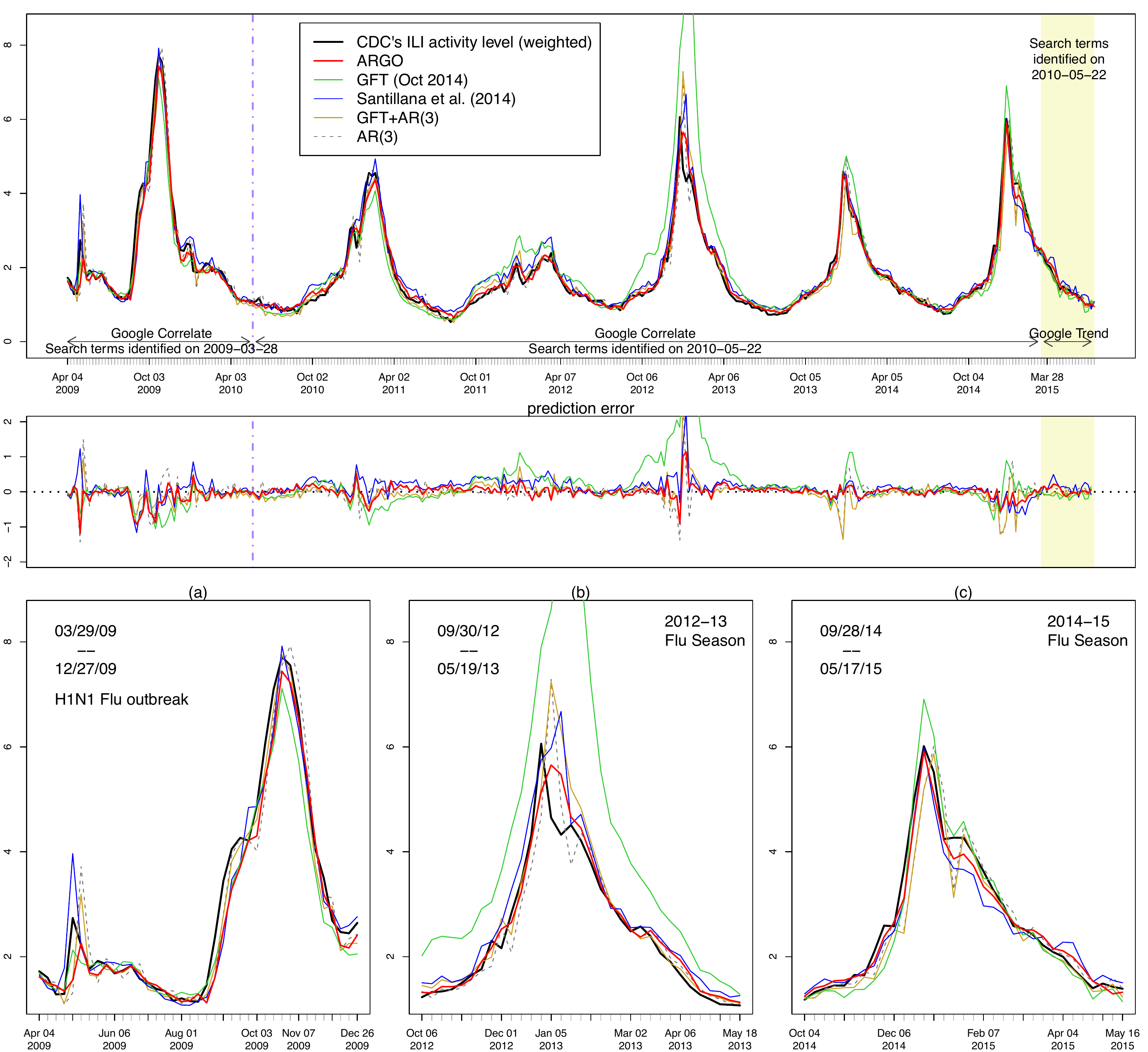}
\end{figure}

Table 1 summarizes these accuracy metrics for all estimation methods for multiple time periods. The first column shows that ARGO's estimates outperform all other alternatives, in every accuracy metric for the whole time period. The other columns of Table 1 show the performance of all the methods for the 2009 off-season H1N1 flu outbreak, and each regular flu season since 2010.
The panels of Figure \ref{fig:all_pred} display the estimates against the observed CDC-reported ILI activity level. 

Close inspection shows that, in the post-2009 regular flu seasons, ARGO uniformly outperformed all other alternative estimation methods in terms 
of {\color{black}root mean squared error, mean absolute error, mean absolute percentage error, and correlation}. 
ARGO avoids the notorious over-shooting problem of GFT, as seen in Figure \ref{fig:all_pred}. 
During the 2009 off-season H1N1 flu outbreak, {\color{black}ARGO had the smallest mean absolute percentage error. 
In terms of root mean squared error and mean absolute error, ARGO (relative RMSE = 0.640, relative MAE = 0.584) 
had the second best performance, under-performing slightly only to GFT+AR(3) model (relative RMSE = 0.580, relative MAE = 0.570).
In terms of correlation, ARGO  (r=98.5\%) had similar performance to (the potentially in-sample data of) GFT (r=98.9\%) \cite{Cook_etal_11} 
and GFT+AR(3) model (r=98.6\%), while outperforming all the other alternatives.}

{\color{black}To assess the statistical significance of the improved prediction power of ARGO,
we constructed a 95\% Confidence 
Interval for the relative efficiency of ARGO compared to other benchmark methods. The Relative Efficiency of method 1 to method 2 is the ratio of the true 
Mean Squared Error of method 2 to that of method 1 \cite{everitt2002cambridge}, which can be estimated by its 
observed value (see eq \eqref{eq:relative_efficiency_poist_est}); its confidence interval can be constructed 
by stationary bootstrap of the error residual time series \cite{politis1994stationary}. 
Table \ref{tab:relative_efficiency} shows that ARGO is estimated to be
at least twice as efficient as any other alternative and the improvement in accuracy is highly statistically significant.}

It is well-known that CDC reports undergo revisions, weeks after their initial publication, that respond to internal consistency checks and lead to more accurate estimates 
of patients with ILI symptoms seeking medical attention. Thus, the available historical CDC information, 
in a given week, is not necessarily as accurate as it will be. We tested the effect of using (potentially inaccurate) 
unrevised information by obtaining the historical unrevised and revised reports, 
and the dates when the reports were revised, from the CDC website for the time period of our study. We used 
only the information that would have been available to us, at the time of estimation, and produced a time series of 
estimates for the whole time period described before. We compared our estimates to all other methods and found that ARGO still 
outperformed them all. Moreover, the values of all five accuracy metrics for ARGO essentially did not change, suggesting a desirable 
robustness to revisions in CDC's ILI activity reports. The results are shown in Table S1 in the Supporting Information.


We faced an additional challenge in producing real-time estimates for the {\color{black}latest portion of the} 2014-2015 flu season. 
At the time of writing this article, the only data available to us for the week of March 28, 2015 and later came from the \textit{Google Trends} website. 
The information from \textit{Google Trends} has even lower quality than from \textit{Google Correlate} and changes every week.
These undesired changes affected the quality of our estimates. In order to assess the stability of ARGO in the presence 
of these variations in the data, we obtained the search frequencies of the same query terms from \textit{Google Trends} website 
on {\color{black}25} different days during the month of April 2015, and produced a set of {\color{black}25} historical estimates using ARGO.
The results of the accuracy metrics associated to these estimates are shown in Table S2 in the Supporting Information. This table 
shows that, despite the observed variation in the \textit{Google Trends} data, ARGO is threefold 
more stable than the method of \cite{Santillana_etal_14}, and still outperforms on average any other method.

\begin{table}[t]
\color{black}
\small
\centering
\scalebox{0.9}{
\begin{tabular}{m{.35cm}|r|l|l|lllll}
\hline
 & & Whole period & Off-season flu & \multicolumn{5}{ |c }{Regular flu seasons (week 40 to week 20 next year)}  \\
  \hline
 & &  & H1N1 & 2010-11 & 2011-12  & 2012-13  & 2013-14  & 2014-15  \\
 \hline

\parbox[t]{2mm}{\multirow{6}{*}{\rotatebox[origin=c]{90}{\textbf{RMSE}}}}&  ARGO  & \textbf{0.608} & 0.640 & \textbf{0.596} & \textbf{0.807} & \textbf{0.687} & \textbf{0.306} & \textbf{0.438} \\ 
&  GFT (Oct 2014)  & 2.216 & 0.773 & 1.110 & 3.023 & 4.451 & 0.986 & 0.700 \\ 
&  Santillana et al. (2014)  & 0.915 & 0.833 & 0.881 & 2.027 & 1.090 & 0.446 & 0.663 \\ 
&  GFT+AR(3)  & 0.912 & \textbf{0.580} & 0.602 & 1.382 & 1.279 & 0.993 & 0.906 \\ 
&  AR(3)  & 0.957 & 0.813 & 0.794 & 1.051 & 1.191 & 0.969 & 0.928 \\ 
&  Naive  & 1 (0.348) & 1 (0.600) & 1 (0.339) & 1 (0.163) & 1 (0.499) & 1 (0.350) & 1 (0.465) \\ 
\hline
\parbox[t]{2mm}{\multirow{6}{*}{\rotatebox[origin=c]{90}{\textbf{MAE}}}}&  ARGO   & \textbf{0.649} & 0.584 & \textbf{0.574} & \textbf{0.748} & \textbf{0.650} & \textbf{0.391} & \textbf{0.530} \\ 
&  GFT (Oct 2014)   & 1.834 & 0.777 & 1.260 & 3.277 & 5.028 & 0.891 & 0.770 \\ 
&  Santillana et al. (2014)   & 1.052 & 0.719 & 1.010 & 2.211 & 1.029 & 0.610 & 0.820 \\ 
&  GFT+AR(3)   & 0.888 & \textbf{0.570} & 0.613 & 1.308 & 1.016 & 1.034 & 0.839 \\ 
&  AR(3)   & 0.925 & 0.777 & 0.787 & 0.951 & 0.988 & 0.917 & 0.934 \\ 
&  Naive   & 1 (0.201) & 1 (0.425) & 1 (0.259) & 1 (0.135) & 1 (0.325) & 1 (0.212) & 1 (0.295) \\ 
\hline
\parbox[t]{2mm}{\multirow{6}{*}{\rotatebox[origin=c]{90}{\textbf{MAPE}}}}&  ARGO    & \textbf{0.787} & \textbf{0.620} & \textbf{0.663} & \textbf{0.770} & \textbf{0.719} & \textbf{0.453} & \textbf{0.620} \\ 
&  GFT (Oct 2014)    & 1.937 & 0.721 & 1.394 & 3.442 & 5.419 & 0.892 & 0.895 \\ 
&  Santillana et al. (2014)    & 1.381 & 0.765 & 1.380 & 2.306 & 1.251 & 0.754 & 0.958 \\ 
&  GFT+AR(3)    & 1.037 & 0.683 & 0.698 & 1.407 & 0.986 & 1.062 & 0.828 \\ 
&  AR(3)    & 1.003 & 0.894 & 0.814 & 0.947 & 0.939 & 0.891 & 0.916 \\ 
&  Naive    & 1 (0.090) & 1 (0.139) & 1 (0.105) & 1 (0.081) & 1 (0.110) & 1 (0.084) & 1 (0.097) \\ 
\hline
\parbox[t]{2mm}{\multirow{6}{*}{\rotatebox[origin=c]{90}{\textbf{Correlation}}}}&  ARGO     & \textbf{0.986} & 0.985 & \textbf{0.989} & \textbf{0.928} & \textbf{0.968} & \textbf{0.993} & \textbf{0.993} \\ 
&  GFT (Oct 2014)     & 0.875 & \textbf{0.989} & 0.968 & 0.833 & 0.926 & 0.969 & 0.986 \\ 
&  Santillana et al. (2014)     & 0.971 & 0.967 & 0.983 & 0.927 & 0.956 & 0.985 & 0.984 \\ 
&  GFT+AR(3)     & 0.967 & 0.986 & 0.985 & 0.879 & 0.929 & 0.945 & 0.957 \\ 
&  AR(3)     & 0.964 & 0.968 & 0.971 & 0.877 & 0.903 & 0.927 & 0.945 \\ 
&  Naive     & 0.961 & 0.951 & 0.954 & 0.887 & 0.924 & 0.923 & 0.937 \\ 
\hline
\parbox[t]{2mm}{\multirow{6}{*}{\rotatebox[origin=c]{90}{\shortstack{\textbf{Corr. of} \\ \textbf{increment}}}}}&  ARGO      & \textbf{0.758} & 0.806 & \textbf{0.810} & 0.286 & \textbf{0.527} & \textbf{0.938} & 0.912 \\ 
&  GFT (Oct 2014)      & 0.706 & \textbf{0.863} & 0.702 & 0.484 & 0.502 & 0.847 & \textbf{0.918} \\ 
&  Santillana et al. (2014)      & 0.690 & 0.776 & 0.693 & \textbf{0.510} & 0.367 & 0.915 & 0.889 \\ 
&  GFT+AR(3)      & 0.512 & 0.708 & 0.708 & 0.165 & 0.141 & 0.534 & 0.587 \\ 
&  AR(3)      & 0.385 & 0.585 & 0.569 & 0.077 & 0.011 & 0.404 & 0.493 \\ 
&  Naive      & 0.436 & 0.602 & 0.570 & 0.095 & 0.134 & 0.406 & 0.514 \\ 
   \hline

\end{tabular}
}
\caption{Comparison of different models for the estimation of influenza epidemics. 
{\color{black} GFT+AR(3) stands for the model $p_t = \mu + \alpha_1 p_{t-1} + \alpha_2 p_{t-2} + \alpha_3 p_{t-3} + \beta \mathrm{GFT}(t)$, where the GFT estimate
is treated as an exogenous variable.} Boldface highlights the 
best performance for each metric in each study period. RMSE, MAE and {\color{black}MAPE} are relative to the error of naive method;
that is, the number reported is the ratio of error of a given method to that of the naive method. The absolute error of the 
naive method is reported in the parentheses. {\color{black}All comparisons are based on the original scale of ILI activity level.}}
\label{tab:results}
\end{table}

\begin{table}[t]
\color{black}
\centering
\begin{tabular}{r|r|r}
  \hline
 & point estimate & 95\% CI \\ 
  \hline
GFT (Oct 2014) & 12.85 & [5.18, 91.82] \\ 
  Santillana et al. (2014) & 2.02 & [1.36, 2.83] \\ 
  GFT+AR(3) & 2.17 & [1.23, 4.53] \\ 
  AR(3) & 2.40 & [1.56, 3.69] \\ 
   \hline
\end{tabular}
\caption{\color{black}Estimate of Relative Efficiency of ARGO compared to other models with 95\% Confidence Interval (CI). Relative Efficiency being 
larger than one suggests increased predictive power of ARGO compared to the alternative method.}\label{tab:relative_efficiency}
\end{table}

\section{Discussion}
\subsection{Strength of ARGO}
The results presented here demonstrate the superiority of our approach both in terms of accuracy and robustness, when compared to all existing flu tracking models based on Google searches. The value of these results is even higher given the fact that they were produced with low quality input variables. It is highly likely that our methodology would lead to even more accurate results if we were given access to the input variables that Google uses to calculate their estimates.

The combination of seasonal flu information with dynamic reweighting of search information, appears to be a key factor 
in the enhanced accuracy of ARGO. The level of ILI activity last week typically has a significant effect on the current 
level of ILI activity, and ILI activity half a year ago and/or one year ago could provide further information, as shown 
in Figure S1 of the Supporting Information, which reflects a strong temporal auto-correlation. The integration of time series information leads 
to a smooth and continuous estimation curve and prevents undesired spikes. {\color{black} However, simply adding GFT to an 
autoregressive model is suboptimal compared to ARGO, because simply treating GFT as an individual variable is incapable to adjust 
for time series information at the resolution of individual query terms, and many terms included in GFT may no longer provide extra information 
once time series information is incorporated.} In fact, once the time series information is included,
fewer Google search query 
terms remain significant. For example, among 100 \textit{Google Correlate} query terms, ARGO 
selected 14 terms on average each week, whereas the method of \cite{Santillana_etal_14} and GFT \cite{Ginsberg_2009} 
{\color{black}selected 38 and 45 terms each week on average, respectively}. The combination of ARGO's smoothness and sparsity lead 
to a substantial reduction on the estimation error, {\color{black} as observed in Tables 1 and 2, where ARGO shows 
improved performance in all evaluation metrics over the whole time period and is twice as efficient as GFT+AR(3).}


Our methodology allows us to transparently understand how Google search information and historical flu information complement one another. 
Time series models tend to be slow in response to sudden observed changes in CDC's ILI activity level. 
The AR(3) model shows this ``delaying'' effect, despite its seemingly good correlation. Google searches, on the other hand, are better at 
detecting sudden ILI activity changes, but are also very sensitive to public's over-reaction.    

To investigate further the responsiveness (co-movement) of ARGO towards the change in ILI activity, we calculated the correlation of increment 
between each estimation model and CDC's ILI activity level. The correlation of increment between two time series $a_t$ and $b_t$ is defined 
as $\mathrm{Corr}(a_t - a_{t-1}, b_t - b_{t-1})$, which measures {\color{black}how well $a_t$ captures the changes in $b_t$}. 
Table 1 shows that ARGO has similar {\color{black}capability in capturing the changes in ILI level 
to that of GFT and the method of \cite{Santillana_etal_14}}, while outperforming the time series model AR(3) uniformly. 


Time series information (seasonality) tends to pull ARGO's estimate towards the historical level. 
This was evident at the onset of the off-season H1N1 flu outbreak (week ending at 05/02/2009), 
which resulted in ARGO's under-estimation. ARGO self-corrected its performance the following week by shifting a portion of model 
weights from the time series domain to the Google searches domain. Inversely, at the height of 2012-13 season, ARGO, GFT and the method 
of \cite{Santillana_etal_14} all missed the peak due to an unprecedented surge of search activity. 
ARGO achieved the fastest self-correction by redistributing the weights not only across Google terms but also across time 
series terms, missing the peak by only 1 week, as opposed to 2 weeks for \cite{Santillana_etal_14} and about 4 weeks for GFT. It is important to note that while we have used CDC's ILI as our gold standard for influenza activity in the US population, 
and data from Google Correlate/Trends as our independent variables, our methodology can be immediately adapted to 
any other suitable ILI gold standard and/or set of independent variables. 

\subsection{Limitations and next steps} 
While ARGO displays a clear superiority over previous methods, it is not fail-proof. Since it relies on the public's search behavior, 
any abrupt changes to the inner works of the search engine or any changes in the way health-related search information is 
displayed to users will affect the accuracy of our methodology \cite{Tsukayama2014,Gianatasio2014}. We expect that ARGO will be 
fast at correcting itself if any such change takes place in the future. 
As in any predictive method, the quality of past performance does not guarantee the quality of future performance.
In this article, we fixed the search query terms after 2010 so as 
to directly compare our results with GFT, which kept the same query terms since 2010; future application of ARGO may update search terms more frequently.
ARGO can be easily generalized to any temporal and spatial scales for a variety of diseases or social events amenable to be tracked by 
internet searches or services \cite{yuan2013monitoring,Paul_etal_14,chan2011using,preis2013, althouse2011prediction, ocampo2013using, Yang2011, Cavazos2014}.
Further improvements in influenza prediction may come from combining multiple predictors constructed from disparate data sources \cite{Santillana2015_ensemble}.
After the submission of this article, Google announced that GFT would be discontinued and that their raw data would be made accessible to selected scientific teams. This announcement happened soon after the GFT team published a manuscript that proposed a new time-series based 
method for the (now discontinued) GFT engine \cite{lampos2015advances}. This new development makes our contribution timely and useful in providing a transparent method for disease tracking in the future. 

\section{Materials and Methods}
{\color{black}All data used in this article are publicly available. Therefore, IRB approval is not needed.}

\subsection{Google Data}
{\color{black} To avoid forward-looking information in our out-of-sample predictions, and to make 
the search term selection in our approach consistent with the main revision to GFT \cite{Cook_etal_11} immediately 
after the H1N1 pandemic, we obtained the highest correlated terms to the CDC's ILI using Google 
correlate ({\color{black}www.google.com/trends/correlate}) for 
two different time periods. For the first time period (pre-H1N1 period), we inserted only CDC's ILI data 
from Jan 2004 to March 28, 2009 into Google Correlate, and used the resulting most highly correlated 
search terms as independent variables for our out-of-sample predictions for the time period 
April 4, 2009 - May 22, 2010. For the second time period (post-H1N1), we inserted only CDC's ILI data 
from Jan 2004 to May 22, 2010 into Google Correlate to select new search terms as done in \cite{Cook_etal_11}. 
These last search terms were used as independent variables for all subsequent predictions presented in this work. 
Tables S4 and S5 in the Supporting Information show all query terms identified. For the pre-H1N1 period (the first time period), the terms from Google Correlate 
include spurious (or over-fitted) terms like ``march vacation'' or ``basketball standings'', 
as discussed in \cite{Lazer_etal_14}. However, Figure S1 in the Supporting Information shows that 
these spurious terms were often not selected by ARGO, i.e., ARGO would give them zero weights, demonstrating its robustness. 
For the post-H1N1 time period, the updated query terms from Google Correlate include mostly flu-related terms (see Table S5 in 
Supporting Information). This suggests that spurious terms were ``filtered-out'' by including off-season flu data. For the time period 
of March 28, 2015 up to the date of submission of this article, 
we acquired search frequencies for this set of query terms from Google Trends ({\color{black}www.google.com/trends. Date of access: July 11, 2015})} 
as Google Correlate only provides data up to March 28, 2015 at the time of writing this article.

\noindent Google Correlate standardizes the search volume of each query to have mean zero and standard deviation one across time and contains 
data only from 2004 to {\color{black}Mar 2015}. 
To make Google Correlate data compatible with Google Trends data, we linearly transformed the Google Correlate data to the same scale of 0 to 100 in our analysis. We used Google Correlate data up to its last available date, and then switched to Google Trends 
data afterwards. This is indicated in Figure \ref{fig:all_pred} by different shades of the background. 
We used the latest version of Google Flu Trends (4th version, revised in Oct 2014) weekly estimates of ILI activity level as one of our comparison methods. GFT is available at www.google.org/flutrends/us/data.txt ({\color{black}Date of access: 2015-07-11}).

\subsection{CDC's data} We use the weighted version of CDC's ILI activity level as the estimation target (available at 
gis.cdc.gov/grasp/fluview/fluportaldashboard.html. {\color{black}Date of access: 2015-07-11}). The weekly revisions of CDC's ILI are available at the CDC website for all recorded seasons (from week 40 of a given year to week 20 of the subsequent year). For example, ILI report revision at week 50 of season 2012-13 is available at 
www.cdc.gov/flu/weekly/weeklyarchives2012-2013/data/senAllregt50.htm; 
ILI report revision at week 9 of season 2014-15 is available at 
www.cdc.gov/flu/weekly/weeklyarchives2014-2015/data/senAllregt09.html.

\subsection{Formulation of our model}
Our model ARGO is motivated by a hidden Markov model. 
The \textit{logit}-transformed CDC-reported ILI activity level 
{\color{black} $\{y_t\}$} is the intrinsic time series of interest. We impose an autoregressive (AR) model with lag $N$ on it, 
which implies that the collection of vectors $\{y_{(t-N+1):t}\}_{t \ge N}$ is a Markov chain {\color{black}(this captures the clinical fact that 
flu lasts for a period, but not indefinitely)}. The vector of \textit{log}-transformed normalized 
volume of Google search queries at time $t$, $\bm X_{t}$, depends only on the ILI activity at the same time, $y_t$ {\color{black}(this follows the intuition that flu 
occurrence causes people to search flu related information online)}. The Markovian property on block $y_{(t-N+1):t}$ leads to the (vector) hidden Markov model structure.
\begin{equation}
\begin{matrix}
y_{1:N} & \rightarrow & y_{2:(N+1)} & \rightarrow & \cdots &  \rightarrow & y_{(T-N+1):T} \\
\downarrow & & \downarrow & &         & &       \downarrow \\
\bm X_N& & \bm X_{N+1} &&&& \bm X_T
\end{matrix}
\end{equation}
Our formal mathematical assumptions are: \\
(1) $y_t = \mu_y + \sum_{j=1}^N \alpha_j y_{t-j} + \epsilon_t,\quad \epsilon_t \overset{iid}{\sim} \mathcal{N}(0,\sigma^2)$ \\
(2) $\bm X_t \mid y_t \sim \mathcal{N}_K\left(\bm\mu_x +  y_t\bm\beta, \bm{Q}\right)$ \\
(3) Conditional on $y_t$, $\bm X_t$ is independent of $\{y_l, \bm X_l: l \neq t\}$ \\
where $\bm\beta = (\beta_1, \beta_2, \ldots,\beta_K)^\intercal, \bm\mu_x = (\mu_{x_1}, \mu_{x_2}, \ldots, \mu_{x_K})^\intercal$, and $\bm{Q}$ is the covariance matrix. To make the variables more normal, we transform the original ILI activity level $p_t$ from $[0, 1]$ to $\mathbb{R}$ using the logit function, 
obtaining the $y_t$, and transform the Google search volumes from $[0,100]$ to $\mathbb{R}$ using the log function, obtaining the $\bm X_t$. 
The log function is appropriate because Google search frequencies usually have exponential growth rate near peaks and are artificially scaled to $[0,100]$ 
by dividing the running maximum. Since \textit{Google Trends} is in integer scale from 0 to 100, we add a small 
number $\delta=0.5$ before the transformation to avoid taking the log of $0$.
The predictive distribution $f\left(y_t\Big| y_{1:(t-1)}, \bm X_{1:t}\right)$ is normal
with mean linear in $y_{(t-N):(t-1)}$ and $\bm X_t$ and constant variance (see the Supporting Information).
This observation leads to equation \eqref{eq:argo} below, which defines the ARGO model.

\subsection{The ARGO model}
Let $y_t=\mathrm{logit}(p_t)$ be the \textit{logit}-transformed CDC's (weighted) ILI activity level $p_t$ at time $t$,
and $X_{i,t}$ the \textit{log}-transfomred Google search frequency of term $i$ at time $t$.
Our ARGO model is given by
\begin{equation}\label{eq:argo}
y_t =  \mu_y + \sum_{j=1}^N \alpha_j y_{t-j} + \sum_{i=1}^K \beta_i X_{i,t} + \epsilon_t,\quad \epsilon_t \overset{iid}{\sim} \mathcal{N}(0, \sigma^2),
\end{equation}
where $\bm X_t$ can be thought as the exogenous variables to time series $\{y_t\}$.

\subsection{Parameter estimation of ARGO model} We chose $N=52$ (weeks) to capture the within-year seasonality in ILI 
activity, and $K=100$ (Google search terms) following the data availability from \textit{Google Correlate}. Since we 
have more independent variables than the number of observations, the usual maximum likelihood estimate 
(ordinary least squares) method will fail. Therefore, we impose regularities for parameter estimation. 
In general we have three kinds of penalties, $L_1$ penalty \cite{tib96}, $L_2$ penalty \cite{ridge70}, and a linear combination of 
$L_1$ and $L_2$ penalties \cite{zouhastie05}. All parameters are dynamically trained every week with a 2-year (104 weeks) rolling window.

\noindent In a given week, the goal is to find parameters $\mu_y$, $\bm \alpha= (\alpha_1, ...,\alpha_{52})$, and $\bm \beta= (\beta_1, ...,\beta_{100})$ that minimize
\begin{align}\label{eq:ar_exogen_est}
\sum_t&\left(y_t - \mu_y - \sum_{j=1}^{52} \alpha_j y_{t-j} - \sum_{i=1}^{100} \beta_i X_{i,t} \right)^2 \notag\\
&+ \lambda_\alpha  \|\bm \alpha\|_1 + \eta_\alpha \|\bm \alpha\|_2^2 + \lambda_\beta  \|\bm \beta\|_1 + \eta_\beta \|\bm \beta\|_2^2
\end{align}
where $\lambda_\alpha,\lambda_\beta, \eta_\alpha, \eta_\beta$ are hyper-parameters.
Ideally, we would like to
use cross-validation to select all 4 hyper-parameters. However, since we have only 104 training data points at a given week due to the two-year moving window, the cross-validation result is highly noisy. Thus, we need to pre-specify some of the hyper-parameters. For model simplicity and sparsity, 
combining with the evidence seen from cross-validation, we set $\eta_\alpha = \eta_\beta = 0$, leading to $L_1$ penalization on both autoregressive and Google search terms.
With the remaining $\lambda_\alpha$ and $\lambda_\beta$, the cross-validation results still have considerable variance. By the same sparsity and simplicity consideration, we further constrained $\lambda_\alpha=\lambda_\beta$. Therefore, the ARGO model we finally propose is equation \eqref{eq:ar_exogen_est} with constraint $\eta_\alpha=\eta_\beta=0$ and $\lambda_\alpha = \lambda_\beta$.
A detailed discussion of our specification of the hyper-parameters is provided in the Supporting Information.

\subsection{Accuracy metrics} 
{\color{black}The Root Mean Squared Error (RMSE), Mean Absolute Error (MAE), and Mean Absolute Percentage Error (MAPE) of 
estimator $\hat p$ to the target ILI activity level $p$ are defined, respectively, as 
$\mathrm{RMSE}(\hat p_t, p_t)=\left({\frac{1}{n}\sum_{t=1}^n (\hat p_t - p_t)^2}\right)^{1/2}$, 
$\mathrm{MAE}(\hat p_t, p_t)=\frac{1}{n}\sum_{t=1}^n |\hat p_t - p_t|$, 
$\mathrm{MAPE}(\hat p_t, p_t)=\frac{1}{n}\sum_{t=1}^n {|\hat p_t - p_t|}/{p_t}$. 
}
The correlation of estimator $\hat p$ to the target ILI activity level $p$ is their sample correlation coefficient. 
The correlation of increment between $\hat p_t$ and $p_t$ is defined as \\
$\text{Corr. of increment}(\hat p_t, p_t) = \mathrm{Corr}(\hat p_t - \hat p_{t-1}, p_t - p_{t-1})$.

\noindent
{\color{black}The Relative Efficiency of estimator $\hat p^{(1)}$ to estimator $\hat p^{(2)}$ is
$
e(\hat p^{(1)}, \hat p^{(2)}) = {{\mathrm{MSE}}_{\text{true}}^{(2)}}/{{\mathrm{MSE}}_{\text{true}}^{(1)}}$, where ${\mathrm{MSE}}_{\text{true}}^{(i)} = \mathbb{E}[(\hat p^{(i)}-p)^2]$, 
which can be estimated by
\begin{equation}\label{eq:relative_efficiency_poist_est}
\textstyle
\hat e\left(\hat p^{(1)}, \hat p^{(2)}\right) = \frac{{\mathrm{MSE}}_{\text{obs}}^{(2)}}{{\mathrm{MSE}}_{\text{obs}}^{(1)}} \quad\text{where}\quad 
{\mathrm{MSE}}_{\text{obs}}^{(i)} = \frac{1}{n}\sum_{t=1}^n \left(\hat p_t^{(i)}-p_t\right)^2.
\end{equation}
The 95\% Confidence Interval can be constructed by time series stationary bootstrap method \cite{politis1994stationary}, where the replicated time 
series of the error residual is generated using geometrically distributed random blocks with mean length 52 (which corresponds to one year). 
We obtain the basic bootstrap confidence interval for $\log\left\{ e\left(\hat p^{(1)}, \hat p^{(2)}\right)\right\}$ and then recover the original 
scale by exponentiation. The non-parametric bootstrap confidence interval takes the autocorrelation and cross-correlation of the errors into account, 
and is insensitive to the mean block length.}
\subsection{Acknowledgments}
{\color{black}S. C. Kou's research is supported in part by NSF grant DMS-1510446.}


\begin{thebibliography}{10}

\bibitem{Ginsberg_2009}
Ginsberg J et~al. (2009) Detecting influenza epidemics using search engine
  query data.
\newblock {\em Nature} 457:1012--1014.

\bibitem{Polgreen_2008}
Polgreen PM, Chen Y, Pennock DM, Nelson FD, Weinstein RA (2008) Using internet
  searches for influenza surveillance.
\newblock {\em Clinical Infectious Diseases} 47(11):1443--1448.

\bibitem{yuan2013monitoring}
Yuan Q et~al. (2013) Monitoring influenza epidemics in {C}hina with search
  query from baidu.
\newblock {\em PloS One} 8(5):e64323.

\bibitem{Paul_etal_14}
Paul MJ, Dredze M, Broniatowski D (2014) Twitter improves influenza
  forecasting.
\newblock {\em PLOS Currents Outbreaks}.

\bibitem{McIver_2014}
McIver DJ, Brownstein JS (2014) Wikipedia usage estimates prevalence of
  influenza-like illness in the {U}nited {S}tates in near real-time.
\newblock {\em PLoS Computational Biology} 10(4):e1003581.

\bibitem{Santillana15112014}
Santillana M, Nsoesie EO, Mekaru SR, Scales D, Brownstein JS (2014) Using
  clinicians’ search query data to monitor influenza epidemics.
\newblock {\em Clinical Infectious Diseases} 59(10):1446--1450.

\bibitem{wesolowski2014}
Wesolowski A et~al. (2014) Commentary: Containing the {Ebola} outbreak--the
  potential and challenge of mobile network data.
\newblock {\em PLOS Currents Outbreaks}.

\bibitem{chan2011using}
Chan EH, Sahai V, Conrad C, Brownstein JS (2011) Using web search query data to
  monitor dengue epidemics: a new model for neglected tropical disease
  surveillance.
\newblock {\em PLoS Neglected Tropical Diseases} 5(5):e1206.

\bibitem{preis2013}
Preis T, Moat HS, Stanley HE (2013) Quantifying trading behavior in financial
  markets using google trends.
\newblock {\em Scientific Reports} 3(1684).

\bibitem{bollen2011}
Bollen J, Mao H, Zeng X (2011) Twitter mood predicts the stock market.
\newblock {\em Journal of Computational Science} 2(1):1--8.

\bibitem{wu2014}
Wu L, Brynjolfsson E (2015) {\em The future of prediction: How Google searches
  foreshadow housing prices and sales}.
\newblock in {\em Economic Analysis of the Digital Economy}, eds.{}
  Avi~Goldfarb SG, Tucker C.
\newblock (University of Chicago Press), pp. 89--118.

\bibitem{Helft_2008}
Helft M (2008) Google uses searches to track flu's spread (The New York Times).

\bibitem{butler2013}
Butler D (2013) When google got flu wrong.
\newblock {\em Nature} 494(7436):155.

\bibitem{Cook_etal_11}
Cook S, Conrad C, Fowlkes AL, Mohebbi MH (2011) Assessing google flu trends
  performance in the {U}nited {S}tates during the 2009 influenza virus a
  ({H1N1}) pandemic.
\newblock {\em PLoS One} 6(8):e23610.

\bibitem{Lazer_etal_14}
Lazer D, Kennedy R, King G, Vespignani A (2014) The parable of google flu:
  Traps in big data analysis.
\newblock {\em Science} 343(6176):1203--1205.

\bibitem{Santillana_etal_14}
Santillana M, Zhang DW, Althouse BM, Ayers JW (2014) What can digital disease
  detection learn from (an external revision to) google flu trends?
\newblock {\em American Journal of Preventive Medicine} 47(3):341--347.

\bibitem{Stefansen_2014}
Stefansen C (2014) Google flu trends gets a brand new engine.
\newblock
  googleresearch.blogspot.com/2014/10/google-flu-trends-gets-brand-new-engine.html.

\bibitem{WHO_14}
{W}{H}{O} (2014) Influenza (seasonal), fact sheet number 211.
\newblock Accessed April, 2015.

\bibitem{Shaman_2012}
Shaman J, Karspeck A (2012) Forecasting seasonal outbreaks of influenza.
\newblock {\em Proceedings of the National Academy of Sciences}
  109(50):20425--20430.

\bibitem{lipsitch_2011}
Lipsitch M, Finelli L, Heffernan RT, Leung GM, Redd SC (2011) Improving the evidence base for decision making during a pandemic: the example of 2009 influenza a/h1n1.
\newblock {\em Biosecurity and bioterrorism} 9(2):89--115.

\bibitem{nsoesie2014systematic}
Nsoesie EO, Brownstein JS, Ramakrishnan N, Marathe MV (2014) A systematic
  review of studies on forecasting the dynamics of influenza outbreaks.
\newblock {\em Influenza and other respiratory viruses} 8(3):309--316.

\bibitem{chretien2014influenza}
Chretien JP, George D, Shaman J, Chitale RA, McKenzie FE (2014) Influenza
  forecasting in human populations: A scoping review.
\newblock {\em PloS One} 9(4):e94130.

\bibitem{Nsoesie_2013}
Nsoesie E, Mararthe M, Brownstein J (2013) Forecasting peaks of seasonal
  influenza epidemics.
\newblock {\em PLoS Currents} 5.

\bibitem{soebiyanto2010modeling}
Soebiyanto RP, Adimi F, Kiang RK (2010) Modeling and predicting seasonal
  influenza transmission in warm regions using climatological parameters.
\newblock {\em PloS One} 5(3):e9450.

\bibitem{shaman2013real}
Shaman J, Karspeck A, Yang W, Tamerius J, Lipsitch M (2013) Real-time influenza
  forecasts during the 2012--2013 season.
\newblock {\em Nature Communications} 4(2837).

\bibitem{Yang03032015}
Yang W, Lipsitch M, Shaman J (2015) Inference of seasonal and pandemic
  influenza transmission dynamics.
\newblock {\em Proceedings of the National Academy of Sciences}
  112(9):2723--2728.

\bibitem{paolotti2014}
Paolotti D et~al. (2014) Web-based participatory surveillance of infectious
  diseases: the influenzanet participatory surveillance experience.
\newblock {\em Clinical Microbiology and Infection} 20(1):17--21.

\bibitem{dalton2009flutracking}
Dalton C et~al. (2009) Flutracking: a weekly australian community online survey
  of influenza-like illness in 2006, 2007 and 2008.
\newblock {\em communicable diseases intelligence quarterly report}
  33(3):316--22.

\bibitem{Smolinski2015}
Smolinski MS et~al. (2015) Flu near you: Crowdsourced symptom reporting
  spanning two influenza seasons.
\newblock {\em American Journal of Public Health} (0) e1-e7.

\bibitem{althouse2011prediction}
Althouse BM, Ng YY, Cummings DA (2011) Prediction of dengue incidence using
  search query surveillance.
\newblock {\em PLoS Neglected Tropical Diseases} 5(8):e1258.

\bibitem{ocampo2013using}
Ocampo AJ, Chunara R, Brownstein JS (2013) Using search queries for malaria
  surveillance, {T}hailand.
\newblock {\em Malaria Journal} 12(1):390.

\bibitem{scarpino2012optimizing}
Scarpino SV, Dimitrov NB, Meyers LA (2012) Optimizing provider recruitment for
 influenza surveillance networks.
\newblock {\em PLoS Computational Biology} 8(4):e1002472.

\bibitem{Davidson2015}
Davidson MW, Haim DA, Radin JM (2015) Using networks to combine ``big data''
  and traditional surveillance to improve influenza predictions.
\newblock {\em Scientific Reports} 5.

\bibitem{Burkom_etal_2007}
Burkom HS, Murphy SP, Shmueli G (2007) Automated time series forecasting for
  biosurveillance.
\newblock {\em Statistics in Medicine} 26(22):4202--4218.

\bibitem{everitt2002cambridge}
Everitt BS, Skrondal A (2002) The cambridge dictionary of statistics.
\newblock {\em Cambridge: Cambridge}.

\bibitem{politis1994stationary}
Politis DN, Romano JP (1994) The stationary bootstrap.
\newblock {\em Journal of the American Statistical association}
  89(428):1303--1313.

\bibitem{Tsukayama2014}
Tsukayama H (2014) Google is testing live-video medical advice (The Washington
  Post).
\newblock Accessed on April 20, 2015.

\bibitem{Gianatasio2014}
Gianatasio D (2014) How this agency cleverly stopped people from googling their
  medical symptoms: The right ads at the right time (Adweek, Online).
\newblock Accessed on April 20, 2015.

\bibitem{Yang2011}
Yang AC, Tsai SJ, Huang NE, Peng CK (2011) Association of internet search
  trends with suicide death in {Taipei} {City}, {Taiwan}, 2004--2009.
\newblock {\em Journal of Affective Disorders} 132(1):179--184.

\bibitem{Cavazos2014}
Cavazos-Rehg PA et~al. (2014) Monitoring of non-cigarette tobacco use using
  google trends.
\newblock {\em Tobacco Control}.

\bibitem{tib96}
Tibshirani R (1996) Regression shrinkage and selection via the lasso.
\newblock {\em Journal of the Royal Statistical Society. Series B
  (Methodological)} 58(1):267--288.

\bibitem{ridge70}
Hoerl AE, Kennard RW (1970) Ridge regression: Biased estimation for
  nonorthogonal problems.
\newblock {\em Technometrics} 12(1):55--67.

\bibitem{zouhastie05}
Zou H, Hastie T (2005) Regularization and variable selection via the elastic
  net.
\newblock {\em Journal of the Royal Statistical Society: Series B (Statistical
  Methodology)} 67(2):301--320.
  
\bibitem{lampos2015advances}
Lampos V, Miller AC, Crossan S, Stefansen C (2015) Advances in nowcasting influenza-like illness rates using search query logs.
\newblock {\em Scientific reports}. 5:12760--

\bibitem{Santillana2015_ensemble}
Santillana M, Nguyen AT, Dredze M, Paul MJ, Nsoesie E, Brownstein JS (2015) Combining Search, Social Media, and Traditional Data Sources to Improve Influenza Surveillance
\newblock {\em PLoS Comput Biol}, 11(10):e1004513


\end{thebibliography}

%




\clearpage
\begin{appendices}
\noindent {\huge \bf Supporting Information}

\section{SI Methods and Robustness Analysis}
Details of our methodology are presented as follows. First, the predictive distribution in the 
formulation of the ARGO model and the corresponding assumptions are described; 
second, the statistical strategy to determine the hyper--parameters of the ARGO model is explained; 
third, the results of two sensitivity analysis aimed at testing the robustness of the ARGO 
methodology--(a) with respect to subsequent revisions of CDC's ILI activity reports, and (b) with respect 
to observed variation of the input variables coming from \textit{Google Trends} data--are presented; 
{\color{black} fourth, the exact search query terms identified by Google Correlate with different 
data access dates are presented; fifth, a heatmap showing the coefficients for the time series
and Google search terms dynamically trained by ARGO is included.}

\section{Predictive distribution in the formulation of ARGO model}
To improve normality for both the input variables and the dependent variables, 
the CDC-reported ILI activity level was \textit{logit}--transformed, and the linearly normalized volume of Google search queries 
were \textit{log}--transformed. To avoid taking the log of 0, we add a small number $\delta = 0.5$ before the log-transformation. 
These transformations led to two sets of variables, the intrinsic (influenza epidemics activity) time series of interest $\{y_{t}\}$, and the (Google search) variable vector $\bm X_{t}$ at time $t$ (that depends only on $y_t$), respectively. Our formal mathematical assumptions are:
\begin{enumerate}
\item $y_t = \mu_y + \sum_{j=1}^N \alpha_j y_{t-j} + \epsilon_t,\quad \epsilon_t \overset{iid}{\sim} \mathcal{N}(0,\sigma^2)$
\item $\bm X_t \mid y_t \sim \mathcal{N}_K\left(\bm\mu_x +  y_t\bm\beta, \bm{Q}\right)$
\item Conditional on $y_t$, $\bm X_t$ is independent of $\{y_l, \bm X_l: l \neq t\}$
\end{enumerate}
where $\bm\beta = (\beta_1, \beta_2, \ldots,\beta_K)^\intercal, 
\bm\mu_x = (\mu_{x_1}, \mu_{x_2}, \ldots, \mu_{x_K})^\intercal$, and $\bm{Q}$ is the covariance matrix. 
The predictive distribution $f\left(y_{t+1}\Big| y_{1:t}, \bm X_{1:(t+1)}\right)$ is given by
\begin{align}
&{\scriptscriptstyle f\left(y_{t+1}\Big| y_{1:t}, \bm X_{1:(t+1)}\right) }\notag\\
&{\scriptscriptstyle \sim \mathcal{N}\Bigg(\left(\frac{1}{\sigma^2}+\bm\beta^\intercal \bm{Q}^{-1}\bm\beta \right)^{-1}\left(\frac{\mu_y + \bm \alpha^\intercal y_{(t-N+1):t}}{\sigma^2} + \bm\beta^\intercal\bm{Q}^{-1}(\bm X_{t+1}-\bm\mu_{x})\right)}, \notag\\
&{\scriptscriptstyle\quad\quad\quad \left(\frac{1}{\sigma^2}+\bm\beta^\intercal \bm{Q}^{-1}\bm\beta \right)^{-1}\Bigg)}\label{eq:hmm_pred}
\end{align}
which is a normal distribution, whose mean is a linear combination of $y_{(t-N):(t-1)}$ and $\bm X_t$, and whose variance is a constant.

\section{Determination of the hyper--parameters for ARGO}\label{sec:hyperpar}
The optimized parameters of the ARGO model, $\mu_y$, $\bm \alpha= (\alpha_1, ...,\alpha_{N})$, 
$\bm \beta= (\beta_1, ...,\beta_{K})$ are obtained by

\begin{align}\label{eq:ar_exogen_est}
\underset{ \mu_y, \bm\alpha, \bm\beta}{\argmin} \quad &\sum_t\left(y_t - \mu_y - \sum_{j=1}^{52} \alpha_j y_{t-j} - \sum_{i=1}^{100} \beta_i X_{i,t} \right)^2 \notag\\
&+ \lambda_\alpha  \|\bm \alpha\|_1 + \eta_\alpha \|\bm \alpha\|_2^2 + \lambda_\beta  \|\bm \beta\|_1 + \eta_\beta \|\bm \beta\|_2^2.
\end{align}
The training period consists of a two--year ($104$ weeks) rolling window that immediately precedes the desired date of estimation. The hyper--parameters are $\lambda_\alpha,\lambda_\beta, \eta_\alpha, \eta_\beta$. We tested the performance of ARGO with the following specifications of hyper--parameters:
\begin{enumerate}
\item Restrict $\eta_\alpha = \eta_\beta = 0$ and $\lambda_\alpha=\lambda_\beta$, cross validate on $\lambda_\alpha$. This is our proposed ARGO with the same $L_1$ penalty for Google search terms and autoregressive lags.\label{item:same_l1}
\item Restrict $\eta_\alpha = \eta_\beta = 0$, cross validate on $(\lambda_\alpha, \lambda_\beta)$. This is ARGO with separate $L_1$ penalties for Google search terms and autoregressive lags.\label{item:sep_l1}
\item Restrict $\eta_\alpha = \eta_\beta $ and $\lambda_\alpha = \lambda_\beta = 0$, cross validate on $\eta_\alpha$.  This is ARGO with the same $L_2$ penalty for Google search terms and autoregressive lags.\label{item:same_l2}
\item Restrict $\lambda_\alpha = \lambda_\beta = 0$, cross validate on $(\eta_\alpha, \eta_\beta)$.  This is ARGO with separate $L_2$ penalties for Google search terms and autoregressive lags.\label{item:sep_l2}
\item Restrict $\lambda_\alpha = \lambda_\beta, \eta_\alpha = \eta_\beta$, cross validate on $(\lambda_\alpha,  \eta_\alpha)$.  This is ARGO with the same elastic net (both $L_1$ and $L_2$) penalty for Google search terms and autoregressive lags.\label{item:enet}
\end{enumerate}

\noindent Table \ref{tab:hyperpar} summarizes the in-sample estimation performance for our proposed ARGO, 
together with the other specifications of hyper--parameters. It is apparent from the table that 
the $L_1$ penalty generally outperforms $L_2$ penalty. The $L_1$ penalty tends to shrink the 
coefficients of unnecessary independent variables to be exactly zero, and thus eliminates redundant information; 
on the other hand, the $L_2$ penalty can only shrink the coefficients to be close to zero. As a 
result, $L_2$ penalized coefficients are not as sparse as their $L_1$ counterparts. 
Furthermore, from Table \ref{tab:hyperpar}, we see that ARGO with separate $L_1$ penalties (Specification \ref{item:sep_l1}) outperforms ARGO with separate $L_2$ penalties (Specification \ref{item:sep_l2}), in terms of both root mean squared error and mean absolute error. Similarly, ARGO with the same $L_1$ penalty (Specification \ref{item:same_l1}) outperforms ARGO with the same $L_2$ penalty (Specification \ref{item:same_l2}), in terms of both root mean squared error and mean absolute error.

\noindent The elastic net model, which combines $L_1$ penalty and $L_2$ penalty, does not provide any error reduction. 
In the cross-validation process of setting $(\lambda_\alpha,  \eta_\alpha)$ for the elastic net model, 
70 weeks out of 116 in-sample weeks showed that the smallest cross-validation mean error when restricting $\eta_\alpha=0$ (i.e. zero $L_2$ penalty) is within one standard deviation of the global smallest cross-validation mean error, suggesting that restricting $L_2$ penalty term to be zero (i.e.  $\eta_\alpha=0$) will introduce little bias. Therefore, for the simplicity and sparsity of the model, we drop the $L_2$ penalty terms and use only $L_1$ penalty.

Next we want to decide between the remaining two specifications, ARGO with separate $L_1$ penalties 
(Specification \ref{item:sep_l1}), and ARGO with the same $L_1$ penalty (Specification \ref{item:same_l1}). 
One might argue that Google search terms and autoregressive lags are different sources of information and thus should have different $L_1$ penalties. However, empirical evidence in Table \ref{tab:hyperpar} shows that, again, giving extra flexibility to $(\lambda_\alpha, \lambda_\beta)$ does not generate improvement compared to fixing $\lambda_\alpha = \lambda_\beta$. In the cross-validation process of setting $(\lambda_\alpha, \lambda_\beta)$ for separate $L_1$ penalties, 99 weeks out of 116 in-sample weeks showed that the smallest cross-validation mean error when restricting $\lambda_\alpha = \lambda_\beta$ (i.e. same $L_1$ penalty) is within one standard deviation of the global smallest cross-validation mean error. This may well be due to the gain from variance reduction when imposing the restriction $\lambda_\alpha = \lambda_\beta$. Based on the same simplicity and sparsity consideration, we finally decided to restrict $\eta_\alpha = \eta_\beta = 0$ and $\lambda_\alpha = \lambda_\beta$ in the setting of hyper--parameters for ARGO.

\section{Revision of CDC's ILI activity reports}
Within a flu season, CDC reports are constantly revised to improve their accuracy as new information is 
incorporated. Thus, CDC's weighted ILI figures displayed in previously published reports 
may change in subsequent weeks. As a consequence, in a given week the available CDC ILI information 
from the most recent weeks may be inaccurate. To test the robustness of ARGO in the presence of these revisions and mimic the real-time tracking in our retrospective predictions, we trained ARGO and all other alternative models based on the following schedule. 

Suppose $z_{i,j}$ is the CDC-reported ILI activity level of week $i$ accessed at week $j$. Since CDC's ILI activity report is typically delayed for one week, on week $j$ the historical ILI activity level data we have is $\{z_{i,j} : i \le j-1\}$. Due to revisions, ILI activity level of week $i$ accessed at different weeks $z_{i,i+1}, z_{i,i+2}, \ldots$ may be different but will converge to a finalized value $z_{i,\infty}$ eventually. Hence, to avoid using forward--looking information, in week $j$, we train all models with the ILI activity level accessed at that week $\{z_{i,j} : i \le j-1\}$. In this sense, any future revision beyond week $j$ will not be incorporated in the training at week $j$. Yet for the accuracy metrics, the estimation target remains the finalized the ILI activity level ($z_{i,\infty}, i =1,2,\ldots$). 


Table \ref{tab:results} shows the estimation results when using the aforementioned schedule.  Note that ARGO still outperforms all other alternative models. Moreover, the absolute values of all four accuracy metrics for ARGO trained this way essentially do not change compared to ARGO trained with finalized ILI activity level in the main text, indicating the robustness of ARGO.

The weekly revisions of CDC's ILI activity reports are available at CDC website from week 40 of the year to week 20 of the subsequent year for all seasons studied in this article. For example, ILI activity level revisions at week 50 of season 2012-2013 are available at 
http://www.cdc.gov/flu/weekly/
weeklyarchives2012-
2013/data/senAllregt50.htm; 
ILI activity report revision at week 9 of season 2014-2015 is available at 
http://www.cdc.gov /flu/weekly/weeklyarchives2014-2015/data/senAllregt09.html (the webpage has suffix ``htm'' for seasons before 2014-2015 and suffix ``html'' for 2014-2015 season). In this retrospective case study, when the revisions of ILI activity level were not available for a particular week during off-season period, the finalized ILI activity level was used instead.

\section{Variations of Google Trends data}
\textit{Google Trends} historical data constantly change as a consequence of re-normalizations and algorithm updates. 
To study the robustness of ARGO to \textit{Google Trends} data revisions, we obtained the search 
frequencies of {\color{black} the search query terms identified by \textit{Google Correlate} on May 22, 2010 
(see Figure 2 in the main text and Table \ref{tab:phrases} below)} from the \textit{Google Trends} 
website (http://www.google.com/trends) on {\color{black}25} different days in {\color{black}April 2015}. 
We studied the variability of ARGO's performance when using these {\color{black} 25} different versions 
of \textit{Google Trends} data as input variables for the common time period of Sep 28, 2014 to Mar 29, 2015. 
We {\color{black}studied} the 2014-15 flu season only partially (up to March 2015) because this is the longest 
study period covered by all the obtained versions of \textit{Google Trends} data, 
{\color{black} at the time (May 1, 2015) of the first submission of this article. We want to emphasize that
Google Correlate data were only available up to Feb 2014 when accessed in April 2015}.



Despite the inevitable variation to the revision of the low-quality data from \textit{Google Trends}, ARGO 
still achieves considerable stability compared to the method of Santillana et al.\ \cite{Santillana_etal_14} 
during this time period. Table \ref{tab:sensi_GT} suggests that ARGO is threefold more robust than the method of 
\cite{Santillana_etal_14}. The incorporation of time series information helps ARGO achieve the stability. 
As an extreme example, AR(3) model focuses entirely on the time series information and 
is thus independent of \textit{Google Trends} data revisions. GFT, formulated with the original search variables 
as inputs, is by construction insensitive to the changes in \textit{Google Trends} data. For this portion 
of the study, we included the signal from GFT for context only and we treat it as exogenous in our analysis. 
Based on the results from previous time periods, it is highly likely that if we had access to Google's internal 
raw data (i.e., historical search volume for disease-related phrases) we would have achieved the same 
stability as well. Yet even with these low-quality data, ARGO outperforms GFT uniformly on all versions of data in terms of both root mean squared error and mean absolute error.





{\color{black}

\section{Detailed description of Google Correlate data} 

Tables \ref{tab:phrases09} and \ref{tab:phrases} list the search query phrases identified by
Google Correlate as of March 28, 2009 and of May 22, 2010, respectively. The March 2009 version
included spurious terms such as ``college.basketball.standings'', ``march.vacation'', ``aloha.ski'', 
``virginia.wrestling'', etc. These spurious terms did not appear in the May 2010 version.
}

\section{Dynamic coefficients for ARGO}

\begin{figure}[b]
\centering
\caption{Dynamic coefficients for ARGO. Red color represents positive coefficients, blue color represents negative coefficients, white color represents zero, and grey color represents missing values. Missing values can be the result of (a) query terms not identified by Google Correlate and (b) Google Trends data not available for particular query terms. Black horizontal dashed line separates Google query queries from autoregressive lags. {\color{black}Yellow vertical dashed line separates coefficients trained on \textit{Google Correlate} data from those trained on \textit{Google Trends} data, and green vertical dashed line separates query terms identified on 2009-03-28 from those identified on 2010-05-22.} }\label{fig:coef}
\includegraphics[scale=0.6, page=1]{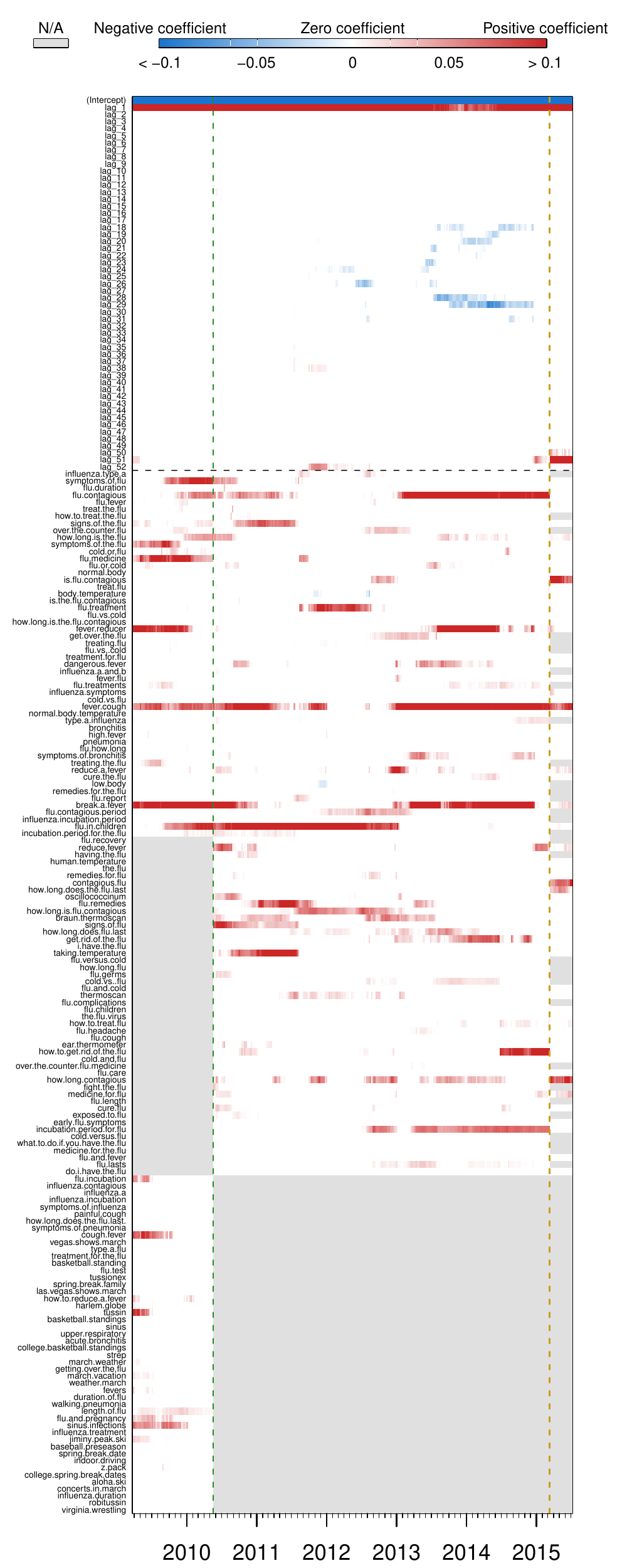}
\end{figure}

Figure \ref{fig:coef} shows the coefficients for the time series
and Google search terms dynamically trained by ARGO via a heatmap. 
The level of ILI activity last week is seen to have a significant effect on the current 
level of ILI activity, and ILI activity half a year ago and/or one year ago could provide further information as the figure shows. Among \textit{Google Correlate} query terms, ARGO 
selected 14 terms out of 100 on average each week.

\clearpage

\begin{table}[ht]
\color{black}
\centering
\scalebox{0.7}{
\begin{tabular}{r|l|l|lllll}
\hline
 & Whole period & Off-season flu & \multicolumn{4}{ |c }{Regular flu seasons (week 40 to week 20 next year)}  \\
  \hline
 &  & H1N1 & 2010-11 & 2011-12  & 2012-13  & 2013-14 & 2014-15 \\
& $\begin{bmatrix}  03/29/09  \\  07/18/15  \end{bmatrix}$ & $\begin{bmatrix} 03/29/09 \\ 12/27/09 \end{bmatrix}$ & $\begin{bmatrix} 10/03/10 \\ 05/22/11 \end{bmatrix}$ & $\begin{bmatrix} 10/02/11 \\ 05/20/12 \end{bmatrix}$ & $\begin{bmatrix} 09/30/12 \\ 05/19/13 \end{bmatrix}$ & $\begin{bmatrix} 09/29/13 \\ 05/18/14 \end{bmatrix}$ & $\begin{bmatrix} 09/28/14 \\ 05/17/15 \end{bmatrix}$ \\
 \hline

\textbf{RMSE}\hfill\vadjust{}  &  &  &  &  &  &  &  \\ 
  ARGO  & \textbf{0.565} & 0.630 & \textbf{0.509} & \textbf{0.608} & \textbf{0.622} & \textbf{0.298} & \textbf{0.434} \\ 
  GFT (Oct 2014)  & 2.003 & 0.702 & 0.971 & 1.878 & 4.387 & 0.885 & 0.714 \\ 
  Santillana et al. (2014)  & 0.897 & 0.858 & 0.760 & 1.179 & 1.248 & 0.373 & 0.691 \\ 
  GFT+AR(3)  & 0.825 & \textbf{0.530} & 0.616 & 0.680 & 1.168 & 0.981 & 0.898 \\ 
  AR(3)  & 0.963 & 0.805 & 0.986 & 1.136 & 1.087 & 0.946 & 0.931 \\ 
  Naive  & 1.000 (0.385) & 1.000 (0.661) & 1.000 (0.388) & 1.000 (0.263) & 1.000 (0.506) & 1.000 (0.391) & 1.000 (0.456) \\ 
  \textbf{MAE}\hfill\vadjust{}   &  &  &  &  &  &  &  \\ 
  ARGO   & \textbf{0.557} & 0.595 & \textbf{0.483} & \textbf{0.555} & \textbf{0.627} & \textbf{0.339} & \textbf{0.501} \\ 
  GFT (Oct 2014)   & 1.465 & 0.670 & 1.093 & 2.026 & 5.082 & 0.747 & 0.787 \\ 
  Santillana et al. (2014)   & 0.865 & 0.723 & 0.875 & 1.283 & 1.087 & 0.472 & 0.847 \\ 
  GFT+AR(3)   & 0.790 & \textbf{0.485} & 0.672 & 0.643 & 1.000 & 1.036 & 0.890 \\ 
  AR(3)   & 0.999 & 0.808 & 0.982 & 1.158 & 1.094 & 0.943 & 0.920 \\ 
  Naive   & 1.000 (0.252) & 1.000 (0.494) & 1.000 (0.299) & 1.000 (0.218) & 1.000 (0.322) & 1.000 (0.253) & 1.000 (0.289) \\ 
  \textbf{MAPE}\hfill\vadjust{}    &  &  &  &  &  &  &  \\ 
  ARGO    & \textbf{0.587} & \textbf{0.587} & \textbf{0.511} & \textbf{0.560} & \textbf{0.588} & \textbf{0.350} & \textbf{0.582} \\ 
  GFT (Oct 2014)    & 1.350 & 0.603 & 1.163 & 2.163 & 4.827 & 0.688 & 0.906 \\ 
  Santillana et al. (2014)    & 0.970 & 0.709 & 1.141 & 1.363 & 1.143 & 0.545 & 0.937 \\ 
  GFT+AR(3)    & 0.848 & 0.599 & 0.749 & 0.669 & 0.819 & 1.068 & 0.964 \\ 
  AR(3)    & 1.067 & 0.915 & 1.051 & 1.169 & 1.050 & 0.945 & 0.935 \\ 
  Naive    & 1.000 (0.129) & 1.000 (0.166) & 1.000 (0.126) & 1.000 (0.129) & 1.000 (0.123) & 1.000 (0.108) & 1.000 (0.095) \\ 
  \textbf{Correlation}\hfill\vadjust{}     &  &  &  &  &  &  &  \\ 
  ARGO     & \textbf{0.985} & 0.979 & \textbf{0.988} & 0.911 & \textbf{0.971} & \textbf{0.992} & \textbf{0.992} \\ 
  GFT (Oct 2014)     & 0.875 & \textbf{0.989} & 0.968 & 0.833 & 0.926 & 0.969 & 0.986 \\ 
  Santillana et al. (2014)     & 0.965 & 0.956 & 0.985 & \textbf{0.937} & 0.938 & 0.987 & 0.973 \\ 
  GFT+AR(3)     & 0.971 & 0.984 & 0.983 & 0.853 & 0.931 & 0.943 & 0.960 \\ 
  AR(3)     & 0.961 & 0.965 & 0.955 & 0.815 & 0.921 & 0.920 & 0.953 \\ 
  Naive     & 0.956 & 0.943 & 0.946 & 0.828 & 0.928 & 0.910 & 0.945 \\ 
  \textbf{Corr. of increment}\hfill\vadjust{}      &  &  &  &  &  &  &  \\ 
  ARGO      & \textbf{0.742} & 0.751 & \textbf{0.772} & 0.262 & \textbf{0.633} & 0.898 & 0.892 \\ 
  GFT (Oct 2014)      & 0.706 & \textbf{0.863} & 0.702 & 0.484 & 0.502 & 0.847 & \textbf{0.918} \\ 
  Santillana et al. (2014)      & 0.625 & 0.680 & 0.719 & \textbf{0.619} & 0.293 & \textbf{0.917} & 0.837 \\ 
  GFT+AR(3)      & 0.536 & 0.703 & 0.703 & 0.155 & 0.220 & 0.514 & 0.621 \\ 
  AR(3)      & 0.420 & 0.562 & 0.554 & 0.067 & 0.106 & 0.360 & 0.549 \\ 
  Naive      & 0.455 & 0.552 & 0.556 & 0.162 & 0.247 & 0.345 & 0.586 \\ 
   \hline

\end{tabular}
}
\caption{Comparison of different models for the estimation of influenza epidemics, with weekly CDC's ILI activity level that excludes forward-looking information from ILI activity report revision. The estimation target is the finalized CDC's ILI activity level. 
RMSE, MAE and {\color{black}MAPE} are relative to the error of naive method. The absolute error of the naive method is reported in the parentheses. \note{\color{black} update table columns, first and last two. ***version date: 2015/04/25***}}

\label{tab:results}
\end{table}

\clearpage

\begin{table}[ht]
\centering
\color{black}
\scalebox{0.9}{
\begin{tabular}{rrrrrr}
  \hline
 & RMSE & MAE & MAPE & Correlation & Corr. of increment \\ 
  \hline
\textbf{Mean} \hfill\vadjust{} &  &  &  &  & \\    
ARGO & 0.226 & 0.304 & 0.079 & 0.981 & 0.831 \\ 
  GFT (Oct 2014) & 0.262 & 0.366 & 0.089 & 0.985 & 0.920 \\ 
  Santillana et al. (2014) & 0.306 & 0.398 & 0.116 & 0.973 & 0.803 \\ 
  GFT+AR(3) & 0.303 & 0.482 & 0.090 & 0.948 & 0.581 \\ 
  AR(3) & 0.332 & 0.492 & 0.096 & 0.936 & 0.492 \\ 
    \hline
\textbf{Standard Deviation} \hfill\vadjust{} &  &  &  &  &\\    
ARGO & 0.013 & 0.017 & 0.005 & 0.002 & 0.016 \\ 
  GFT (Oct 2014) & 0.000 & 0.000 & 0.000 & 0.000 & 0.000 \\ 
  Santillana et al. (2014) & 0.029 & 0.049 & 0.013 & 0.005 & 0.050 \\ 
  GFT+AR(3) & 0.000 & 0.000 & 0.000 & 0.000 & 0.000 \\ 
  AR(3) & 0.000 & 0.000 & 0.000 & 0.000 & 0.000 \\ 
   \hline
\end{tabular}
}
\caption{Mean and Standard Deviation of accuracy metrics when using \textit{Google Trends} data accessed at 
different dates. The common study period is 2014-15 partial season (Sep 28, 2014 to Mar 29, 2015). 
{\color{black}At the time of first submitting this article, Google Correlate data covered only upto Feb 2014, 
which inspired us to study the robustness of ARGO with respect to Google Trends data variability on 
the 2014-15 season.} \note{\color{black} version date: 2015/04/25}}
\label{tab:sensi_GT}
\end{table}

\clearpage
\begin{table}[ht]
\centering
\scalebox{0.9}{
\begin{tabular}{rllll}
  \hline
 & Whole in-sample period & 2006-07 partial season & 2007-08 season & 2008-09 partial season \\
  & 01/07/07-03/29/09 & 01/07/07-05/20/07 & 09/30/07-05/18/08 & 09/28/08-03/29/09 \\
  \hline
\textbf{RMSE}
\hfill\vadjust{}  &  &  &  &  \\
  ARGO w/ same $L_1$ & 0.644 & 0.697 & 0.602 & 0.653 \\
  ARGO w/ sep. $L_1$ & 0.658 & 0.672 & 0.637 & 0.629 \\
  ARGO w/ same $L_2$ & 1.165 & 0.817 & 1.175 & 1.243 \\
  ARGO w/ sep. $L_2$ & 1.010 & 0.740 & 0.946 & 1.173 \\
  ARGO w/ ElasticNet & 0.669 & 0.757 & 0.585 & 0.766 \\
  Naive & 1.000 (0.316) & 1.000 (0.286) & 1.000 (0.473) & 1.000 (0.304) \\
       \hline

\textbf{MAE}
\hfill\vadjust{}  &  &  &  &  \\
  ARGO w/ same $L_1$ & 0.678 & 0.651 & 0.584 & 0.634 \\
  ARGO w/ sep. $L_1$ & 0.691 & 0.671 & 0.621 & 0.593 \\
  ARGO w/ same $L_2$ & 1.223 & 0.836 & 1.094 & 1.469 \\
  ARGO w/ sep. $L_2$ & 1.149 & 0.753 & 0.943 & 1.401 \\
  ARGO w/ ElasticNet & 0.738 & 0.718 & 0.613 & 0.780 \\
  Naive & 1.000 (0.206) & 1.000 (0.245) & 1.000 (0.335) & 1.000 (0.226) \\
       \hline

\textbf{Correlation}
\hfill\vadjust{}
&  &  &  &  \\
  ARGO w/ same $L_1$ & 0.987 & 0.977 & 0.983 & 0.977 \\
  ARGO w/ sep. $L_1$ & 0.986 & 0.980 & 0.980 & 0.976 \\
  ARGO w/ same $L_2$ & 0.969 & 0.984 & 0.976 & 0.955 \\
  ARGO w/ sep. $L_2$ & 0.979 & 0.987 & 0.983 & 0.967 \\
  ARGO w/ ElasticNet & 0.987 & 0.984 & 0.986 & 0.975 \\
  Naive & 0.965 & 0.949 & 0.950 & 0.935 \\
         \hline

  \textbf{Corr. of increment}
\hfill\vadjust{}  &  &  &  &  \\
  ARGO w/ same $L_1$ & 0.779 & 0.643 & 0.857 & 0.646 \\
  ARGO w/ sep. $L_1$ & 0.708 & 0.545 & 0.758 & 0.697 \\
  ARGO w/ same $L_2$ & 0.828 & 0.793 & 0.864 & 0.799 \\
  ARGO w/ sep. $L_2$ & 0.845 & 0.795 & 0.881 & 0.824 \\
  ARGO w/ ElasticNet & 0.814 & 0.835 & 0.852 & 0.738 \\
  Naive & 0.623 & 0.473 & 0.756 & 0.322 \\
       \hline
\end{tabular}
}
\caption{Comparison of different specifications of hyper--parameters for in-sample study period. ``ARGO w/ same $L_1$'' is ARGO with the same $L_1$ penalty for Google search terms and autoregressive lags (Specification \ref{item:same_l1}). ``ARGO w/ sep. $L_1$'' is ARGO with separate $L_1$ penalties for Google search terms and autoregressive lags (Specification \ref{item:sep_l1}). ``ARGO w/ same $L_2$'' is ARGO with the same $L_2$ penalty for Google search terms and autoregressive lags (Specification \ref{item:same_l2}).  ``ARGO w/ sep. $L_2$'' is ARGO with separate $L_2$ penalties for Google search terms and autoregressive lags (Specification \ref{item:sep_l2}). ``ARGO w/ ElasticNet'' is ARGO with the same elastic net penalty for Google search terms and autoregressive lags (Specification \ref{item:enet}). The first column is for the entire in-sample study period. The second column is for 2006-07 partial season. 2006-07 full season is not available because data prior to Jan 2007 is used for training. The third column is for 2007-08 full season. The fourth column is for 2008-09 partial season. 2008-09 full season is not available because our out-of-sample study period starts in Apr 2009. RMSE and MAE are relative to the error of naive method. The absolute error of the naive method is reported in the parentheses. 
}
\label{tab:hyperpar}
\end{table}


\clearpage
\begin{table}[ht]
\centering
\color{black}
\scalebox{0.9}{
\begin{tabular}{llll}
  \hline
  \hline
influenza.type.a & painful.cough & treatment.for.the.flu & weather.march \\ 
  flu.incubation & fever.flu & basketball.standing & fevers \\ 
  bronchitis & over.the.counter.flu & flu.test & duration.of.flu \\ 
  influenza.contagious & pneumonia & tussionex & flu.contagious.period \\ 
  flu.fever & how.long.is.the.flu & reduce.a.fever & cold.vs.flu \\ 
  influenza.a & flu.how.long & how.long.is.the.flu.contagious & cure.the.flu \\ 
  influenza.incubation & treatment.for.flu & treat.flu & walking.pneumonia \\ 
  flu.contagious & fever.cough & spring.break.family & flu.vs..cold \\ 
  treating.the.flu & flu.medicine & las.vegas.shows.march & length.of.flu \\ 
  type.a.influenza & dangerous.fever & how.to.reduce.a.fever & influenza.a.and.b \\ 
  symptoms.of.the.flu & high.fever & flu.or.cold & flu.and.pregnancy \\ 
  influenza.symptoms & is.flu.contagious & incubation.period.for.the.flu & sinus.infections \\ 
  flu.duration & normal.body & harlem.globe & influenza.treatment \\ 
  flu.report & normal.body.temperature & tussin & jiminy.peak.ski \\ 
  symptoms.of.flu & how.long.does.the.flu.last. & basketball.standings & baseball.preseason \\ 
  influenza.incubation.period & symptoms.of.pneumonia & sinus & spring.break.date \\ 
  how.to.treat.the.flu & signs.of.the.flu & upper.respiratory & indoor.driving \\ 
  treat.the.flu & flu.vs.cold & get.over.the.flu & z.pack \\ 
  symptoms.of.bronchitis & low.body & acute.bronchitis & college.spring.break.dates \\ 
  flu.treatment & cough.fever & body.temperature & aloha.ski \\ 
  symptoms.of.influenza & vegas.shows.march & college.basketball.standings & concerts.in.march \\ 
  treating.flu & is.the.flu.contagious & strep & break.a.fever \\ 
  flu.in.children & type.a.flu & march.weather & influenza.duration \\ 
  fever.reducer & flu.treatments & getting.over.the.flu & robitussin \\ 
  cold.or.flu & remedies.for.the.flu & march.vacation & virginia.wrestling \\ 
   \hline
\end{tabular}
}
\caption{All search phrases identified by Google Correlate using data as of 2009-03-28.}\label{tab:phrases09}
\end{table}

\clearpage

\begin{table}[ht]
\centering
\color{black}
\scalebox{0.9}{
\begin{tabular}{llll}
  \hline
  \hline
influenza.type.a & get.over.the.flu & type.a.influenza & flu.care \\ 
  symptoms.of.flu & treating.flu & i.have.the.flu & how.long.contagious \\ 
  flu.duration & flu.vs..cold & taking.temperature & fight.the.flu \\ 
  flu.contagious & having.the.flu & flu.versus.cold & reduce.a.fever \\ 
  flu.fever & treatment.for.flu & bronchitis & cure.the.flu \\ 
  treat.the.flu & human.temperature & how.long.flu & medicine.for.flu \\ 
  how.to.treat.the.flu & dangerous.fever & flu.germs & flu.length \\ 
  signs.of.the.flu & the.flu & cold.vs..flu & cure.flu \\ 
  over.the.counter.flu & remedies.for.flu & flu.and.cold & exposed.to.flu \\ 
  how.long.is.the.flu & influenza.a.and.b & thermoscan & low.body \\ 
  symptoms.of.the.flu & contagious.flu & flu.complications & early.flu.symptoms \\ 
  flu.recovery & how.long.does.the.flu.last & high.fever & remedies.for.the.flu \\ 
  cold.or.flu & fever.flu & flu.children & flu.report \\ 
  flu.medicine & oscillococcinum & the.flu.virus & incubation.period.for.flu \\ 
  flu.or.cold & flu.remedies & how.to.treat.flu & break.a.fever \\ 
  normal.body & how.long.is.flu.contagious & pneumonia & flu.contagious.period \\ 
  is.flu.contagious & flu.treatments & flu.headache & influenza.incubation.period \\ 
  treat.flu & influenza.symptoms & flu.cough & cold.versus.flu \\ 
  body.temperature & cold.vs.flu & ear.thermometer & flu.in.children \\ 
  is.the.flu.contagious & braun.thermoscan & how.to.get.rid.of.the.flu & what.to.do.if.you.have.the.flu \\ 
  reduce.fever & fever.cough & flu.how.long & medicine.for.the.flu \\ 
  flu.treatment & signs.of.flu & symptoms.of.bronchitis & flu.and.fever \\ 
  flu.vs.cold & how.long.does.flu.last & cold.and.flu & flu.lasts \\ 
  how.long.is.the.flu.contagious & normal.body.temperature & over.the.counter.flu.medicine & incubation.period.for.the.flu \\ 
  fever.reducer & get.rid.of.the.flu & treating.the.flu & do.i.have.the.flu \\ 
   \hline
\end{tabular}
}
\caption{All search phrases identified by Google Correlate using data as of 2010-05-22.}\label{tab:phrases}
\end{table}

\clearpage

\end{appendices}

\end{document}